\documentclass[usenatbib]{mnras}
\usepackage{amsmath}
\usepackage{amsfonts}
\usepackage{amssymb}
\usepackage{graphicx}
\begin{document}

\pagerange{\pageref{firstpage}--\pageref{lastpage}} \pubyear{2017}

\title[Spots on WASP-107 and pulsations of WASP-118]{Starspots on WASP-107 and pulsations of WASP-118}

\author[T.~Mo\v{c}nik et al.]{T.~Mo\v{c}nik,\thanks{E-mail: t.mocnik@keele.ac.uk} C.~Hellier, D.~R.~Anderson, B.~J.~M.~Clark and J.~Southworth\\
Astrophysics Group, Keele University, Staffordshire, ST5 5BG, UK}

\date{Accepted XXX. Received YYY; in original form ZZZ}

\maketitle

\label{firstpage}

\begin{abstract}
By analysing the \textit{K2} short-cadence photometry we detect starspot occultation events in the lightcurve of WASP-107, the host star of a warm-Saturn exoplanet. WASP-107 also shows a rotational modulation with a period of $17.5\pm1.4$\thinspace d. Given that the rotational period is nearly three times the planet's orbital period, one would expect in an aligned system to see starspot occultation events to recur every three transits. The absence of such occultation recurrences suggests a misaligned orbit unless the starspots' lifetimes are shorter than the star's rotational period. We also find stellar variability resembling $\gamma$\thinspace Doradus pulsations in the lightcurve of WASP-118, which hosts an inflated hot Jupiter. The variability is multi-periodic with a variable semi-amplitude of $\sim$\thinspace 200\thinspace ppm. In addition to these findings we use the \textit{K2} data to refine the parameters of both systems, and report non-detections of transit-timing variations, secondary eclipses and any additional transiting planets. We used the upper limits on the secondary-eclipse depths to estimate upper limits on the planetary geometric albedos of 0.7 for WASP-107b and 0.2 for WASP-118b.
\end{abstract}


\section{INTRODUCTION}

The \textit{Kepler} \citep{Borucki10} and \textit{K2} \citep{Howell14} missions have provided the community with high-precision photometric observations of 2449 confirmed transiting exoplanets to date among a total of 2716 confirmed transiting exoplanets, according to NASA Exoplanet Archive\footnote{http://exoplanetarchive.ipac.caltech.edu/}.  In addition, \textit{K2} is also observing exoplanets previously found by the ground-based transit surveys such as WASP \citep{Pollacco06}.  WASP-107 and WASP-118 are among the brightest systems observed by \textit{K2}, which allows for detailed characterisation owing to high-precision lightcurves combined with existing spectroscopic observations performed by \citet{Anderson17} and \citet{Hay16}.

WASP-107b is a warm Saturn in a 5.7-d orbit around a \textit{V}=11.6, K6 main-sequence star \citep{Anderson17}. The planet lies in the transition region between ice giants and gas giants, with a mass of 2.2\thinspace $M_{\rm Nep}$ or 0.12\thinspace $M_{\rm Jup}$, but an inflated radius of 0.94\thinspace $R_{\rm Jup}$. The WASP discovery photometry revealed a possible stellar rotational modulation with a period of $\sim$\thinspace 17\thinspace d and an amplitude of 0.4 per cent. This led \citet{Anderson17} to propose that the host star is magnetically active.

WASP-118b is an inflated hot Jupiter with a mass of 0.51\thinspace $M_{\rm Jup}$ and a radius of 1.4\thinspace $R_{\rm Jup}$. It orbits a \textit{V}=11.0 F6IV/V star every 4.0\thinspace d \citep{Hay16}.

If a transiting planet crosses a starspot it will produce a temporary brightening in the transit \citep{Silva03}. Starspot occultation events can provide an accurate measurement of the stellar rotational period and the obliquity, i.e. the angle between the stellar rotational axis and the planet's orbital axis. Obliquities can tell us about a planet's dynamical history and about planet migration mechanisms. If the obliquity is small, the same starspot can be occulted recurrently at different stellar longitudes, such as in the case of an aligned Qatar-2 system \citep{Mocnik16c,Dai17}. Alternatively, if a system is misaligned, the transit chord will cross stellar active latitudes only at certain preferential phases, such as in the case of HAT-P-11b \citep{Sanchis11}.

The presence of a massive close-in planet may induce tides that can lead to multi-periodic non-radial pulsations, and in special cases radial pulsations of the host star \citep{Schuh10,Herrero11}. However, only a handful set of pulsating exoplanet hosts have been found so far. V391b, for example, has been found orbiting an sdB star through pulsation-timing variations \citep{Silvotti07}. The first main-sequence star where asteroseismology was applied to exoplanetary research is $\mu$\thinspace Arae \citep{Bazot05}. High precision and long-term \textit{Kepler} observations have lead to the discovery of several transiting exoplanet host stars which exhibited solar-like oscillations \citep{Davies16}, whose typical semi-amplitudes are of the order of a few ppm for main-sequence stars \citep{Baudin11}.

WASP-33 was the first transiting-exoplanet host star exhibiting $\delta$\thinspace Scuti pulsations, with a semi-amplitude of about 900\thinspace ppm \citep{CollierCameron10b,Herrero11}. One of the harmonics has a frequency 26 times the orbital frequency, suggesting that WASP-33's pulsations might be induced by the planet. A more confident claim for the planet-induced stellar pulsations was made recently for HAT-P-2 system, a possible low-amplitude $\delta$\thinspace Scuti pulsator with a 87 minutes pulsation period and an amplitude of 40 ppm \citep{deWit17}. Its pulsation modes correspond to exact harmonics of the planet's orbital frequency and are thought to be induced by the transient tidal interactions with its massive (8\thinspace $M_{\rm Jup}$), short-period (5.6\thinspace d) and highly eccentric (\textit{e} = 0.5) planet \citep{deWit17}. Systems such as WASP-33 and HAT-P-2 provide a laboratory to study star--planet interactions. The observations provided by the \textit{K2} mission have led to the discovery of another pulsating transiting exoplanet host star, namely HAT-P-56 which is likely a $\gamma$\thinspace Doradus pulsator \citep{Huang15}. While the exact asteroseismic analysis approach can vary depending on the class of variability, one can investigate any potential star--planet interactions and derive precise stellar mass, radius and the depth-dependant chemical composition, given the appropriate modelling capabilities \citep{Schuh10}.

In this paper, we present a detection of starspots in the \textit{K2} lightcurve of WASP-107, and pulsations in the lightcurve of WASP-118. We also refine system parameters for both systems, search for transit-timing variations, phase-curve modulations and any additional transiting planets, and provide a measurement of the rotational period for WASP-107.

Simultaneously with our paper, \citet{Dai17b} have announced an analysis of the same \textit{K2} short-cadence observations of WASP-107. Their results are in a good agreement with ours.

\section{THE K2 OBSERVATIONS}

WASP-107 was observed by \textit{K2} in the 1-min short-cadence observing mode during Campaign 10 between 2016 July 6 and 2016 September 20. The dominant systematics present in the \textit{K2} lightcurves are the sawtooth-shaped artefacts caused by the drift of the spacecraft. We attempted to correct for the drift artefacts using the same self-flat-fielding (SFF) procedure as described in \citet{Mocnik16a}. However, since the drifts of the spacecraft were variable during this observing campaign, we obtained better results using the K2 Systematic Correction pipeline (K2SC; \citealt{Aigrain16}), which uses break-points to isolate different drift behaviours to overcome the issue of inconsistent drifts. The K2SC procedure was designed for the \textit{K2} long-cadence observing mode and had to be modified slightly to accept the short-cadence data. The main modification was to split the input short-cadence lightcurve into smaller overlapping sections, which were processed individually by K2SC and then merged back together using the overlapping sections. However, since the WASP-107 lightcurve exhibits pronounced modulations, the K2SC failed to correct the drift artefacts properly at three different 2-d-long sections centred at BJD 2457624.7, 2457631.8 and 2457638.3. We replaced these sections with the lightcurve we obtained using the above-mentioned SFF procedure. Using this approach we achieved a median 1-min photometric precision of 260\thinspace ppm, compared to 870\thinspace ppm before the artefact correction, and 300\thinspace ppm using only the SFF procedure. The final drift-corrected lightcurve of WASP-107 is shown in Fig.~1. We used this lightcurve to study the rotational modulation of the host star in Section~6. For all other analyses we used the normalized version of the lightcurve, which we produced with PyKE tool {\scriptsize{KEPFLATTEN}} \citep{Still12}, which divides the measured flux with a low-order polynomial fit with window and step sizes of 3 and 0.3\thinspace d, respectively, which effectively removed any low-frequency modulations.

\begin{figure}
\includegraphics[width=8.3cm]{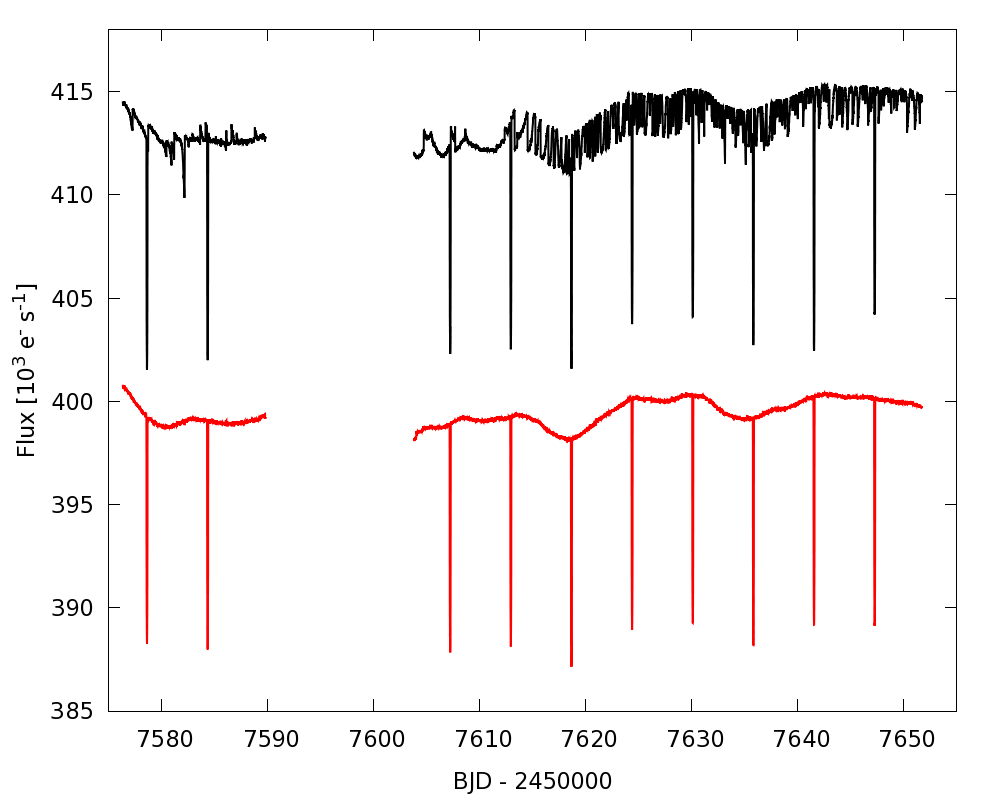}
\caption{The binned lightcurve of WASP-107 shown before (in black) and after the drift correction (red). A 10-min binning was used to reduce the white noise and to display the drift artefacts more clearly. Note the presence of 10 transits and a rotational modulation. The 14-d data gap in the first half of the observing campaign was caused by a failure of CCD module 4, which temporarily powered off the entire photometer. The corrected lightcurve is shown with an offset of \mbox{$-14\thinspace 000\thinspace\rm{e}^{-}\rm{s}^{-1}$} for clarity.}
\end{figure}

WASP-118 was observed during the \textit{K2} Campaign 8 between 2016 January 4 and 2016 March 23, also in the short-cadence mode. Because the spacecraft's drifts were more consistent during Campaign 8, we produced a slightly better lightcurve with our SFF procedure than with K2SC. The drift-corrected lightcurve of WASP-118 is shown in red in Fig.~2. In addition to the transits, the lightcurve exhibits variability with a $\sim$\thinspace 5-d time-scale. The variability is inconsistent and incoherent, and cannot be interpreted as a rotational modulation with confidence. We produced the normalized version of the lightcurve in the same way as for WASP-107. The normalized lightcurve revealed multi-periodic higher-frequency variability with a semi-amplitude of $\sim$\thinspace 200\thinspace ppm (see Fig.~3). We suggest that this variability is produced by the weakly pulsating host star (see Section~7).

\begin{figure}
\includegraphics[width=8.3cm]{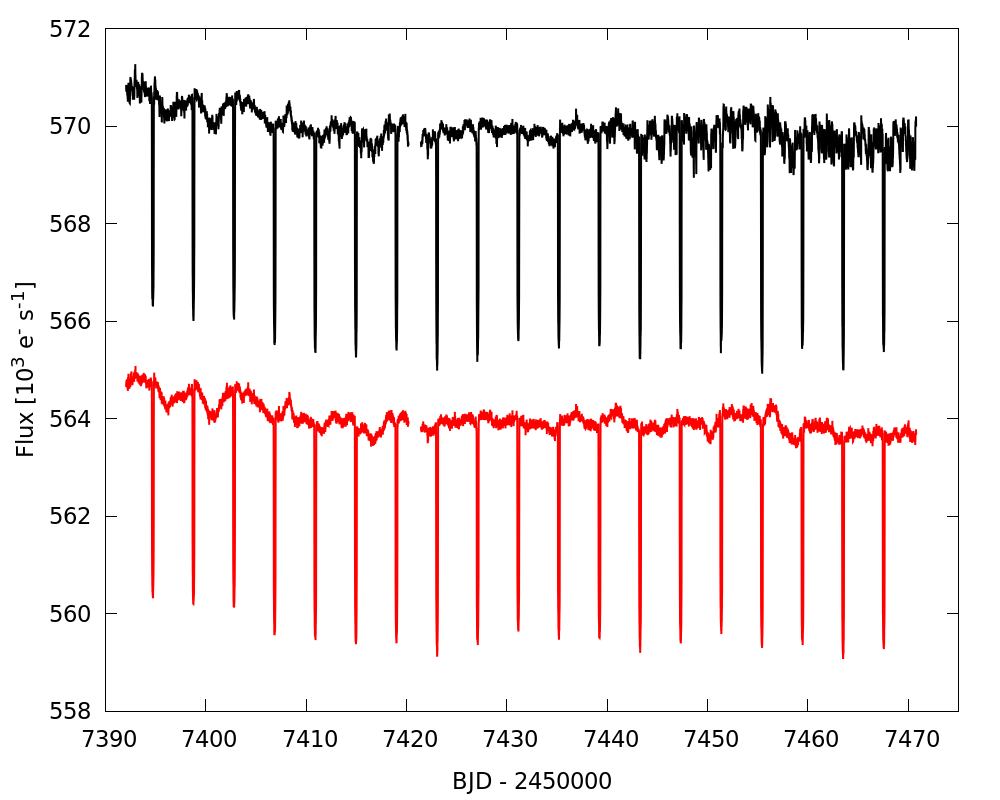}
\caption{The binned lightcurve of WASP-118 shown before (in black) and after the drift correction (red). As in Fig.~1, we applied a 10-min binning to reduce the white noise. 19 transits are visible along with other variability. The corrected lightcurve is shown with an offset of \mbox{$-6\thinspace 000\thinspace\rm{e}^{-}\rm{s}^{-1}$}.}
\end{figure}

\begin{figure*}
\includegraphics[width=17.6cm]{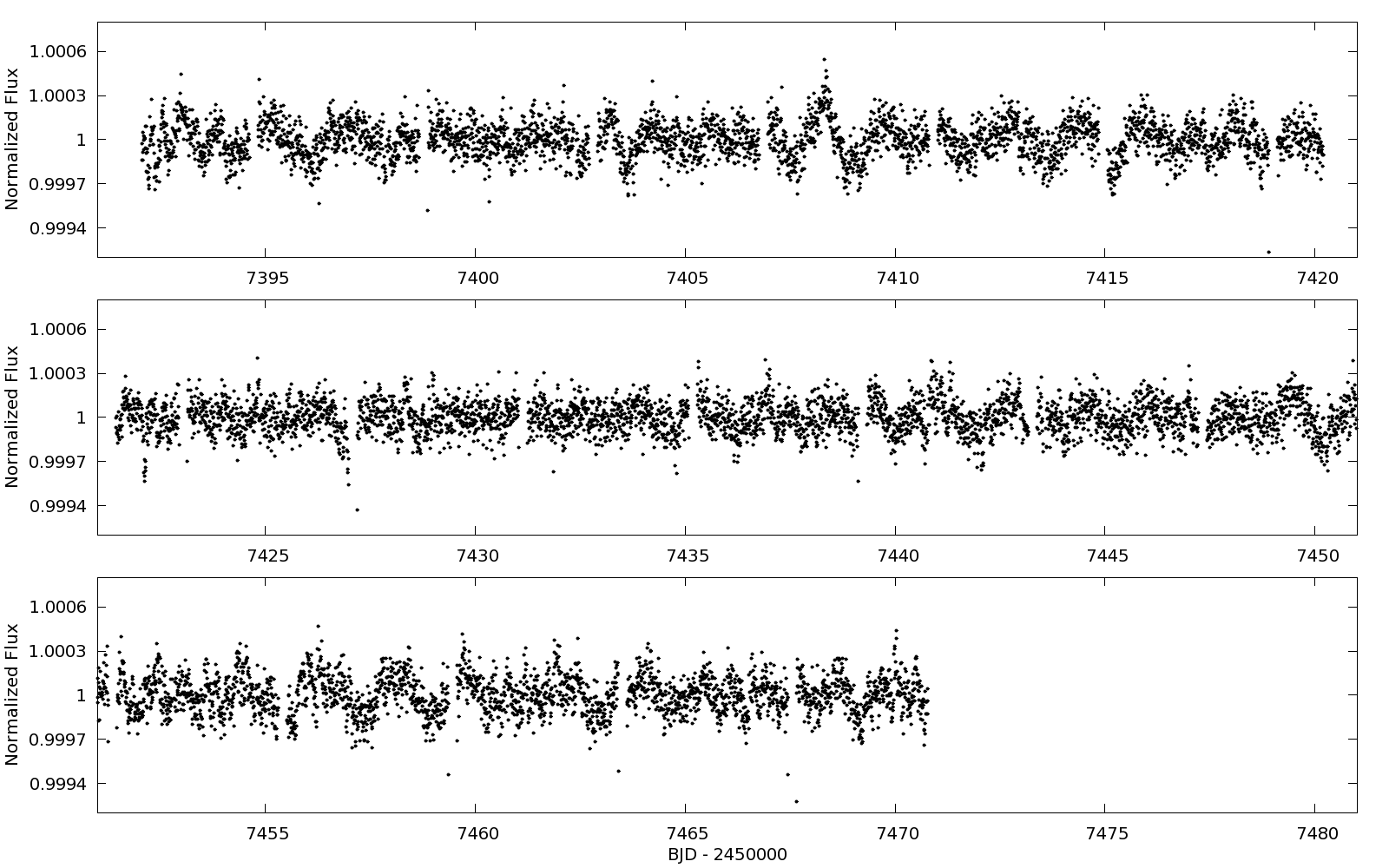}
\caption{Stellar variability of WASP-118 resembling weak $\gamma$\thinspace Doradus pulsations. We show the normalized lightcurve with 10-min binning.}
\end{figure*}

Because the presence of pulsations in the lightcurve of WASP-118 could affect the analysis, we used the lightcurve with embedded pulsations only for the pulsations analysis in Section~7. For every other aspect of analysis we produced another lightcurve in which we removed these higher-frequency, low-amplitude pulsations. This was done by modifying our SFF procedure. Instead of the usual 5-d time-steps, we split the lightcurve into sections of individual spacecraft drifts before applying the SFF correction. Since the typical 6-h time-scale at which the drifts occur is considerably shorter than the observed pulsation time-scales, our modified SFF procedure effectively removed the pulsations from the lightcurve, along with any other variabilities at longer time-scales. This produced the lightcurve with a final median 1-min photometric precision of 210\thinspace ppm. We used this version of lightcurve for every aspect of the analysis, except for the analysis of pulsations in Section 7.

\section{SYSTEM PARAMETERS}

To determine the system parameters we simultaneously analysed the \textit{K2} transit photometry and radial-velocity measurements for both planets using a Markov Chain Monte Carlo (MCMC) code \citep{CollierCameron07,Pollacco08,Anderson15}. We used all the radial-velocity datasets that have been reported in the corresponding discovery papers (\citealt{Anderson17} and \citealt{Hay16}), except the HARPS-N in-transit dataset for WASP-118. This dataset was excluded in order to prevent biasing the system parameters owing to a large scatter and underestimated error bars, as pointed out by the authors of the discovery paper. It may be that some of this scatter results from the pulsations that we now report. Limb darkening was accounted for using a four-parameter law, with coefficients calculated for the \textit{Kepler} bandpass and interpolated from the tabulations of \citet{Sing10}.

For the MCMC analysis of WASP-107 we used the normalized lightcurve, from which we excluded any starspot occultation events (see Section~5), because failing to do so may lead to inaccurate system parameter determination, as discussed by \citet{Oshagh13}. Similarly, the presence of pulsations in the lightcurve of WASP-118 may have affected the precise transit analysis and so we instead used the lightcurve with pulsations removed.

We first imposed a circular orbit in the main MCMC analysis for both planets. We then estimated the upper limits on eccentricities in a separate MCMC run by allowing the eccentricities to be fitted as a free parameter. To improve the precision of the orbital ephemerides we also ran another MCMC analysis for both systems that included all the available ground-based photometric datasets used in the discovery papers which expanded the observational time-span and reduced the uncertainty on the orbital period by factors of 6.5 for WASP-107 and 2.3 for WASP-118. For these additional photometric datasets, we used limb-darkening coefficients from \citet{Claret00,Claret04}, as appropriate for different bandpasses.

The resulting system parameters are given in Table~1 and the corresponding transit models are shown in Figs.~4 and 5.

\begin{figure}
\includegraphics[width=8.3cm]{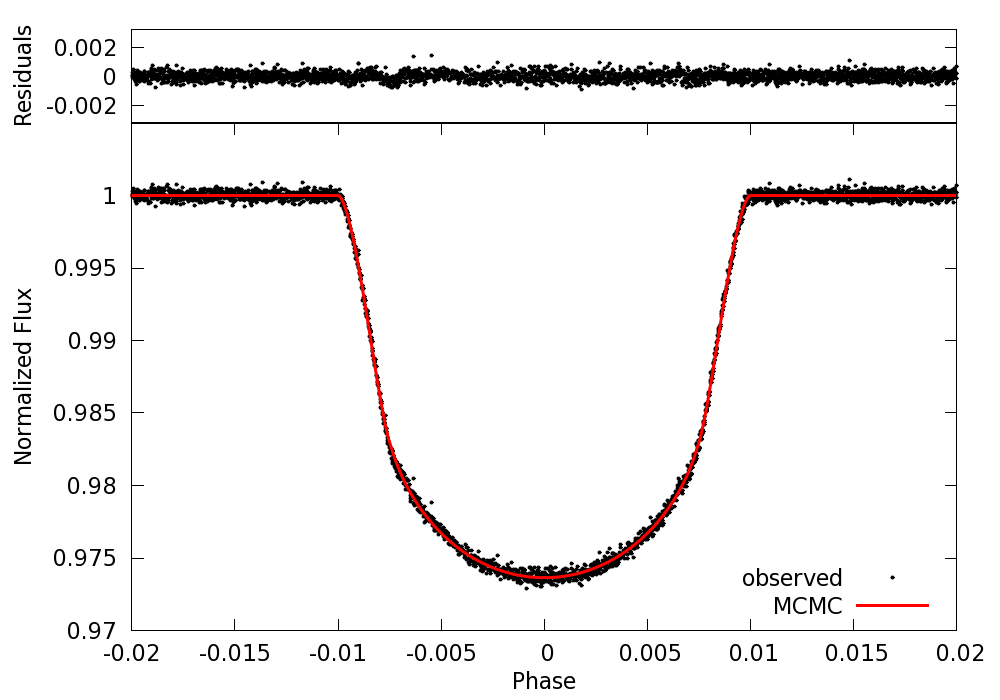}
\caption{Best-fitting MCMC transit model and its residuals for WASP-107.}
\end{figure}

\begin{figure}
\includegraphics[width=8.3cm]{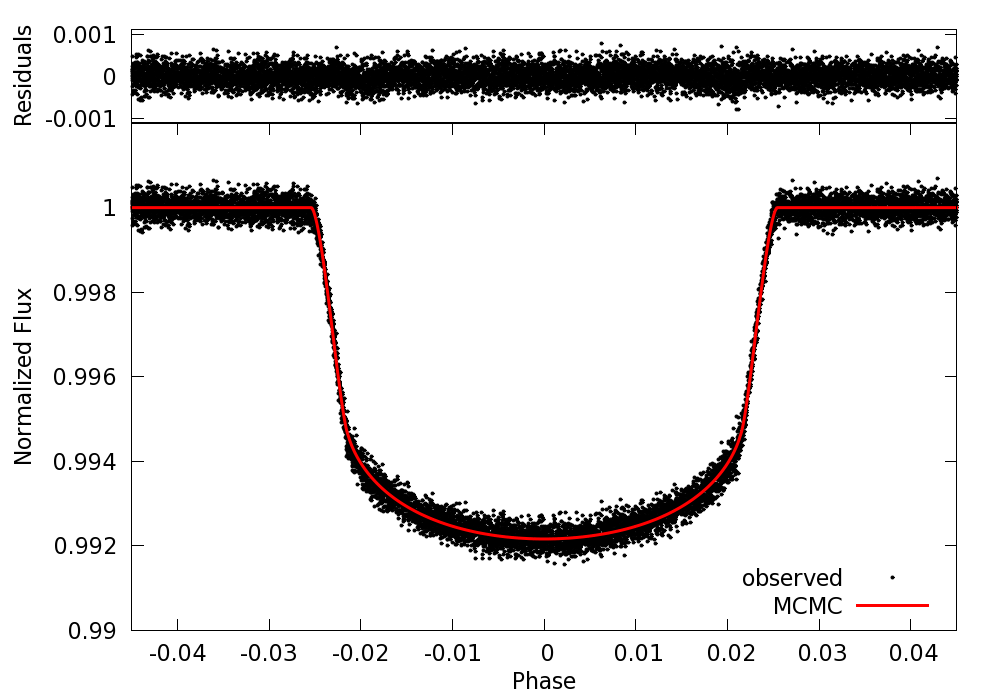}
\caption{Best-fitting MCMC transit model and its residuals for WASP-118.}
\end{figure}

\begin{table*}
\centering
\begin{minipage}{16cm}
\caption{MCMC system parameters for WASP-107 and WASP-118.}
\begin{tabular}{lccccc}
\hline
&&WASP-107&WASP-118&\\
Parameter&Symbol&Value&Value&Unit\\
\hline
Transit epoch&\textit{t}$_{\rm 0}$&$2457584.329746\pm0.000011$&$2457423.044825\pm0.000020$&BJD\\
Orbital period&\textit{P}&$5.72149242\pm0.00000046$&$4.0460407\pm0.0000026$&d\\
Area ratio&$(R_{\rm p}/R_{\star})^{2}$&$0.020910\pm0.000058$&$0.006679\pm0.000010$&...\\
Transit width&\textit{t}$_{\rm 14}$&$0.11411\pm0.000081$&$0.20464\pm0.00010$&d\\
Ingress and egress duration&\textit{t}$_{\rm 12}$, \textit{t}$_{\rm 34}$&$0.01467\pm0.00012$&$0.01610\pm0.00012$&d\\
Impact parameter&\textit{b}&$0.139\pm0.024$&$0.206\pm0.015$&...\\
Orbital inclination&\textit{i}&$89.560\pm0.078$&$88.24\pm0.14$&$^{\circ}$\\
Orbital eccentricity&\textit{e}&0 (adopted; $<$0.025 at $2\sigma$)&0 (adopted; $<$0.028 at $2\sigma$)&...\\
Orbital separation&\textit{a}&$0.0553\pm0.0013$&$0.05450\pm0.00049$&au\\
Stellar mass&\textit{M}$_{\star}$&$0.691\pm0.050$&$1.319\pm0.035$&M$_\odot$\\
Stellar radius&\textit{R}$_{\star}$&$0.657\pm0.016$&$1.754\pm0.016$&R$_\odot$\\
Stellar density&$\rho_{\star}$&$2.441\pm0.023$&$0.2445\pm0.0024$&$\rho_\odot$\\
Planet mass&\textit{M}$_{\rm p}$&$0.119\pm0.014$&$0.52\pm0.18$&$M_{\rm Jup}$\\
Planet radius&\textit{R}$_{\rm p}$&$0.924\pm0.022$&$1.394\pm0.013$&$R_{\rm Jup}$\\
Planet density&$\rho_{\rm p}$&$0.152\pm0.017$&$0.193\pm0.066$&$\rho_{\rm Jup}$\\
Planet equilibrium temperature$^{a}$&\textit{T}$_{\rm p}$&$736\pm17$&$1753\pm34$&K\\
Limb-darkening coefficients&$a_{\rm 1}$, $a_{\rm 2}$, $a_{\rm 3}$, $a_{\rm 4}$&0.710, --0.773, 1.520, --0.641&0.522, 0.313, --0.071, --0.045&...\\
\hline
\end{tabular}
\begin{description}
\setlength\itemsep{0cm}
\item[$^{a}$]Planet equilibrium temperature is based on assumptions of zero Bond albedo and complete heat redistribution.
\end{description}
\end{minipage}
\end{table*}

\section{NO TTV OR TDV}

Inter-planet gravitational interactions can cause transit-timing variations (TTVs) and transit-duration variations (TDVs) \citep{Algol05}. The detection of these variations can therefore reveal additional planets in the system. Typical reported TTV amplitudes range from a few seconds and up to several hours with periods of the order of a few days \citep{Mazeh13}. TDVs are expected to be in phase with the TTVs but at a significantly lower amplitude \citep{Nesvorny13}.

To search for TTVs and TDVs we ran another MCMC analysis on individual transits for both systems. We again removed the starspot occultation events from the WASP-107 lightcurve and used the pulsation-free lightcurve of WASP-118 since the lightcurve variability could affect the timing accuracy \citep{Oshagh13}.

Against the hypothesis of equal transit timing spacings and constant transit durations the measured TTVs and TDVs for WASP-107 correspond to $\chi^{2}$ values of 11.1 and 6.6, respectively, for 10 degrees of freedom. Similarly, for WASP-118 the TTV and TDV $\chi^{2}$ values are 26.8 and 15.7, respectively, for 19 degrees of freedom. Thus there are no significant TTVs or TDVs. The upper limits for WASP-107 are 20 and 60\thinspace s for TTVs and TDVs respectively, for periods shorter than 80\thinspace d. For WASP-118 the upper limits are 40 and 100\thinspace s. Given the absence of any statistically significant TTV or TDV variations, we can conclude that any additional close-in, massive planets are unlikely in either of the two systems.

\section{STARSPOTS ON WASP-107}

In Fig.~6 we show the lightcurve of WASP-107 centred at individual transits after subtracting the best-fitting MCMC transit model from Section~3. The residual lightcurve reveals several starspot occultation events. Each of us has individually examined the residual lightcurve by eye and marked the events as definite or possible occultations. Here we report occultation events that were marked by at least two colleagues. This gives 5 definite starspot occultation events (marked with dark-red ellipses in Fig.~6) and 4 possible events (marked with light-red ellipses) within the 10 observed transits.  Table~2 lists all the marked definite and possible occultation events, measured orbital phases at which they occur and the corresponding stellar longitudes which we calculated using the system parameters from Table~1. There may well be additional, smaller-amplitude spots present in some of the transits, in addition to those listed.

\begin{figure}
\includegraphics[width=8.3cm]{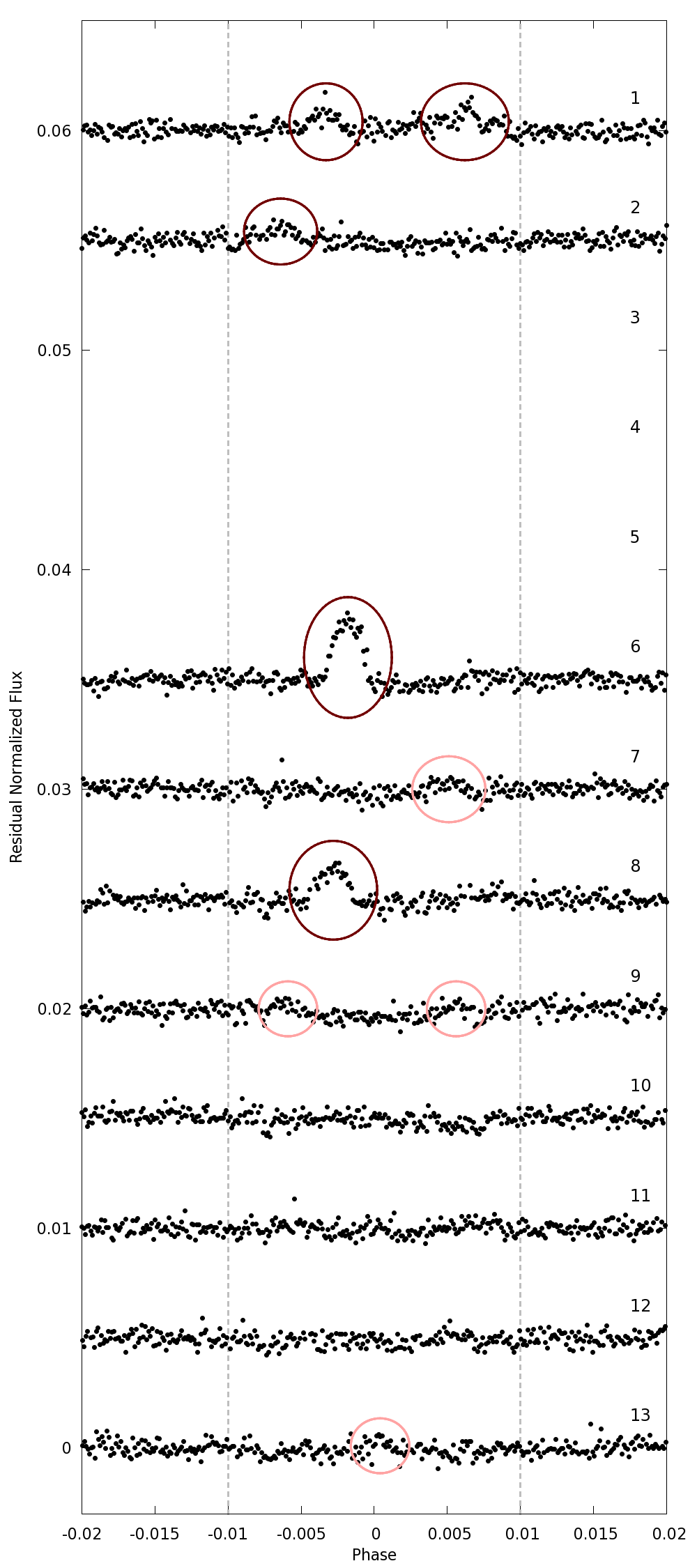}
\caption{Starspot occultations in the model-subtracted lightcurve of WASP-107. Vertical dashed lines show the extent of the transit. Dark-red ellipses mark definite starspot occultation events and light-red ellipses mark possible starspots. Transits 3--5 were not observed owing to the 14-d data gap.}
\end{figure}

\begin{table}
\centering
\caption{Phase and longitude positions of every detected starspot occultation event.}
\begin{tabular}{cr@{\thinspace $\pm$\thinspace}lr@{\thinspace $\pm$\thinspace}l}
\hline
\multicolumn{1}{c}{Transit}&\multicolumn{2}{c}{Phase}&\multicolumn{2}{c}{Stellar}\\
\multicolumn{1}{c}{number}&\multicolumn{2}{c}{}&\multicolumn{2}{c}{longitude$^a$ [$^{\circ}$]}\\
\hline
1&$-0.0033$&$0.0002$&$-22.4$&$1.6$\\
1&$0.0062$&$0.0002$&$45.0$&$2.2$\\
2&$-0.0064$&$0.0002$&$-47.6$&$2.4$\\
6&$-0.0018$&$0.0002$&$-12.1$&$1.4$\\
7&$0.0051$&$0.0002$&$36.2$&$1.9$\\
8&$-0.0028$&$0.0002$&$-18.5$&$1.5$\\
9&$-0.0059$&$0.0002$&$-42.9$&$2.1$\\
9&$0.0056$&$0.0002$&$40.3$&$2.0$\\
13&$0.0004$&$0.0002$&$2.4$&$1.3$\\
\hline
\end{tabular}
\begin{description}
\item[$^{a}$]Longitude runs from $-90^{\circ}$ (first planetary contact), through $0^{\circ}$ (central meridian) to $90^{\circ}$ (last contact).
\end{description}
\end{table}

Given the stellar rotational period of $17.5\pm1.4$\thinspace d (see Section~6) and the planet's orbital period of 5.72\thinspace d (see Table~1), one would expect in an aligned system to see the same starspot being occulted again three transits later, with a longitude shift of $-7^{\circ +31}_{\,\,\, -26}$. Occultations one, two or four transits later would be hard to detect, since the phase shifts would be in multiples of $120^{\circ}$, and therefore the spots would either be close to the limb, where they are hard to detect owing to limb darkening, or not on the visible face at all. 

There is only one pair of starspot occultations that might be a recurrence. If the occultation event in transit 6 and the first occultation in transit 9 (see Fig.~6) were caused by the same starspot, they would imply an orbital period of $18.8\pm0.2$\thinspace d. Although this is compatible with the  $17.5\pm1.4$-d rotational period derived from the rotational modulation (see Section~6) there are reasons to doubt that the pair was actually caused by the same starspot. Firstly, a starspot from transit 8 does not produce an occultation pair at a similar phase shift in transit 11. Secondly, the starspot lifetimes for main-sequence stars are of the order of days (\citealt{Bradshaw14} and citations therein), which means that in 17\thinspace d, the time it takes for the planet to orbit its host star three times, a starspot could disappear.

Overall, we do not find compelling evidence for recurring starspots, which would suggest that the system might be misaligned. However, because of the 17-d time span between recurrences that could be readily observed, we cannot exclude the possibility that it is an aligned system with relatively short starspot lifetimes.

\section{ROTATIONAL MODULATION OF WASP-107}

The lightcurve of WASP-107 in Fig.~1 reveals a low-frequency modulation with a semi-amplitude of about 0.2 per cent. To measure the periodicity of this modulation we removed the planetary transits and calculated a Lomb-Scargle periodogram (see Fig.~7). The highest peak in the periodogram and its full width at half maximum correspond to a periodicity of $17.5\pm1.4$\thinspace d.

\begin{figure}
\includegraphics[width=8.3cm]{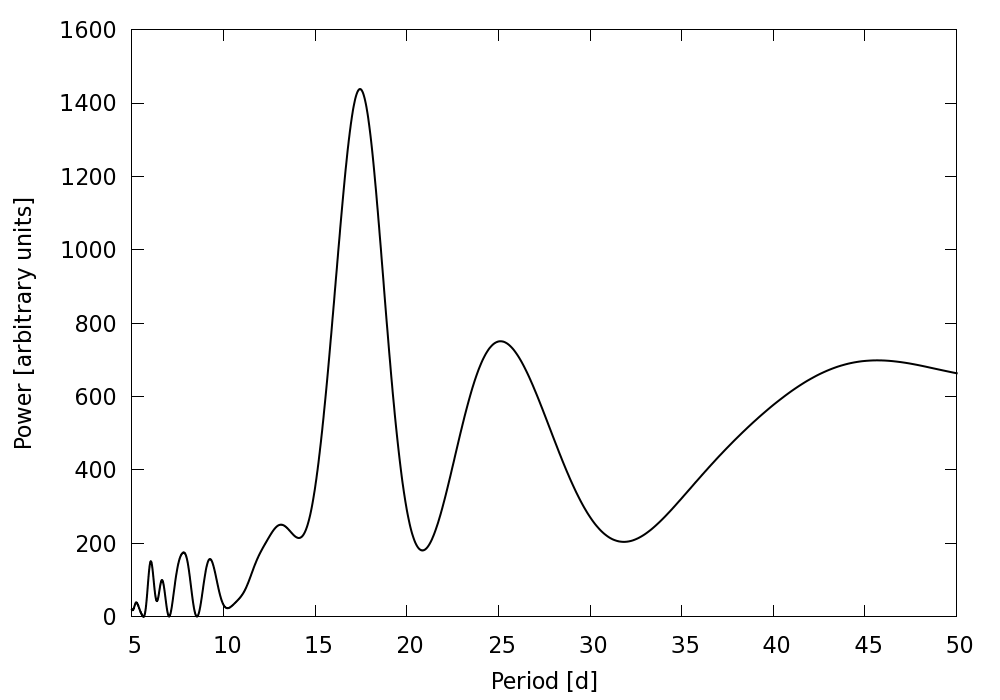}
\caption{Lomb-Scargle periodogram of WASP-107's rotational modulation. The highest peak implies a rotational period of $17.5\pm1.4$\thinspace d.}
\end{figure}

Knowing that the stellar surface harbours starspots (see Section~5), we believe that this modulation is caused by the stellar rotation. Using the stellar radius from Table~1 and assuming that the rotational axis is orthogonal to the line of sight, the measured rotational modulation period implies a stellar rotational velocity of $1.9\pm0.2$\thinspace km\thinspace s$^{-1}$. This value agrees with the spectroscopic projected rotational velocity of $2.5\pm0.8$\thinspace km\thinspace s$^{-1}$ \citep{Anderson17}. Our rotational modulation period and amplitude also agree well with the period of $17\pm1$\thinspace d and amplitude of 0.4 per cent that were derived from the ground-based photometry and reported in the discovery paper by \citet{Anderson17}.

\section{STELLAR PULSATIONS OF WASP-118}

Fig.~3 provides a close-up view of the higher-frequency photometric variability in the lightcurve of WASP-118, after removing incoherent low-frequency modulations. This variability is not correlated with the spacecraft's drifts, is preserved using four different data-reduction procedures, is not present in the \textit{K2} lightcurves of other nearby stars, and is not correlated with the orbital phase. 

The variability cannot be realistically considered as a rotational modulation because its short 2-d period would require the star to rotate three times faster than the spectroscopically measured projected rotational velocity. This would only be possible if the system were hugely misaligned, in contradiction with the Rossiter--McLaughlin measurements by \citet{Hay16} who suggested that the system is aligned.  The variability is therefore most likely to be weak pulsations of the host star.

The semi-amplitude of the pulsations is $\sim$\thinspace 200\thinspace ppm. The Lomb-Scargle periodogram reveals that the variability is multi-periodic (see Fig.~8) with the highest peak at 1.9\thinspace d. To check whether the normalization procedure affected the detected pulsations we ran signal injection tests and found that the applied normalization procedure preserves 80 per cent of the variability at the peak pulsation period near 1.9\thinspace d.

\begin{figure}
\includegraphics[width=8.3cm]{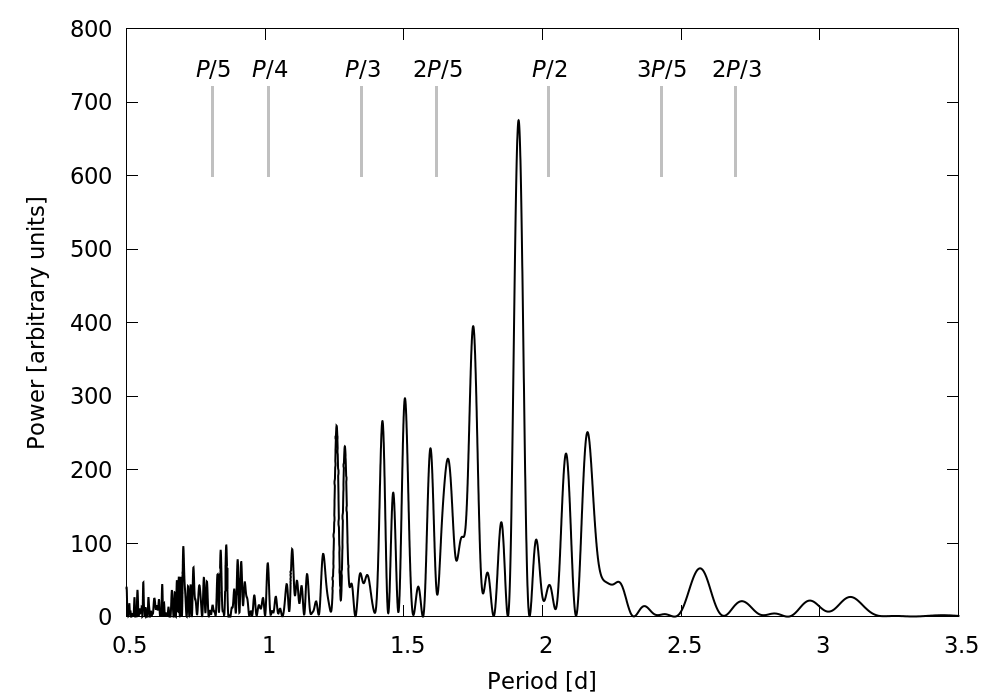}
\caption{The Lomb--Scargle periodogram of WASP-118's higher-frequency variability in the normalized version of the lightcurve. Several peaks between 1 and 2.5\thinspace d indicate that the higher-frequency variability is multi-periodic. Also shown are some of the main harmonics of the planet's orbital period, which demonstrate a mismatch with the pulsation periodicities. Note that  we show here only a few of the main harmonics for clarity. The power spectral density for periods longer than 3.5\thinspace d is virtually zero because of the applied flattening normalization procedure.}
\end{figure}

\citet{Kaye99} introduced a $\gamma$\thinspace Doradus pulsating type for main-sequence and subgiant stars of spectral types A7--F5. $\gamma$\thinspace Doradus stars exhibit non-radial, high-order and low-degree multiperiodic pulsations with periods of 0.4--3\thinspace d and amplitudes below 0.1\thinspace mag. The pulsation amplitude of such stars may vary during an observing season by as much as a factor of 4.

The spectral type and variability characteristics of WASP-118 suggest that the star is probably a weak, late-type $\gamma$\thinspace Doradus pulsator.

If the stellar pulsations were induced by the star--planet interactions, we would expect them to appear at exact harmonics of the planet's orbital period, such as in the cases of WASP-33 \citep{Herrero11} and HAT-P-2 \citep{deWit17}. We found no match when comparing the main harmonics of the WASP-118b's orbital period with the measured pulsations' periodicities (see Fig.~8). Despite the lack of any commensurabilities with the main harmonics, we cannot yet reject any complex commensurabilities nor the possibility that the pulsations were induced by the orbiting planet. A detailed follow-up asteroseismic analysis is required to further investigate the star--planet interactions as a possible cause for the observed pulsations in the lightcurve of WASP-118.

\section{NO PHASE-CURVE MODULATIONS}

Phase-curves in exoplanet systems consist of three main components at optical wavelengths: 1) ellipsoidal modulation, 2) Doppler beaming, and 3) planetary reflection (e.g. \citealt{Esteves13}). Additionally, a transiting planet may produce a secondary eclipse, an occultation of the planet by its host star, which blocks the reflected light during the occultation. Therefore, the depth of the secondary eclipse is twice the semi-amplitude of the reflectional modulation.

To produce a phase-curve of WASP-107 we had to remove the significant rotational modulation prior to phase-folding the lightcurve. However, because the ratio between the 5.7-d orbital period and the 17-d rotational modulation is not small enough, this would also remove any phase-curve modulations. For WASP-118 we again removed stellar variability which would again remove any phase-curve modulations. However, in both systems the procedures would not have removed any secondary eclipses present. A non-detection of secondary eclipses in both systems therefore allowed us to estimate conservative upper limits on the  secondary-eclipse depths of 100\thinspace ppm for WASP-107 and 50\thinspace ppm for WASP-118.

Using the system parameters from Table~1, the theoretically expected semi-amplitudes of ellipsoidal, Doppler beaming and reflectional modulation for WASP-107 are 0.03, 0.2 and $75A_{\rm g}$\thinspace ppm, respectively, where $A_{\rm g}$ is the planet's geometrical albedo. For WASP-118, the expected semi-amplitudes are 1, 0.7 and $150A_{\rm g}$\thinspace ppm. These amplitudes have been calculated using the relations from \citet{Mazeh10}. While the expected amplitudes for the ellipsoidal and Doppler beaming modulations are below the \textit{K2} photometric precision, the inflated planetary radii could produce a significant reflectional modulation in both systems. Using the upper limits for secondary eclipse depths and theoretically expected amplitudes for reflectional modulations allows us to constrain the planetary geometric albedos to less than 0.7 for WASP-107b and less than 0.2 for WASP-118b.

\section{NO ADDITIONAL TRANSITING PLANETS}

To search for signatures of any additional transiting planets in the normalized lightcurves of both systems, we first removed the transits of the known transiting planets and then calculated the box-least-square periodograms of any other periodic signals with the PyKE tool {\scriptsize{KEPBLS}}. The absence of any significant residual signals in the period range 0.5--30\thinspace d results in transit-depth upper limits of 130\thinspace ppm for any additional transits in the WASP-107 system and 140\thinspace ppm in WASP-118.

\section{AGES OF THE HOST STARS}

We estimated the ages of both host stars by comparing the measured stellar densities from Table~1 and the published spectroscopic effective temperatures to isochrones computed from the stellar evolution models. This was done with the Bayesian mass and age estimator {\scriptsize{BAGEMASS}} \citep{Maxted15}, which uses the {\scriptsize{GARSTEC}} code \citep{Weiss08} to compute the evolution models. The best-fitting stellar evolution tracks provided the age estimates of $8.3\pm4.3$\thinspace Gyr for WASP-107 and $2.3\pm0.5$\thinspace Gyr for WASP-118.

The rate at which a star rotates acts as another age estimator. Over time, stars lose angular momentum through magnetised stellar winds and gradually slow down \citep{Barnes03}. Knowing the rotational period of WASP-107 from the detected rotational modulation (see Section~6), we estimated the stellar age with the gyrochronological relation by \citet{Barnes07} to obtain $0.6\pm0.2$\thinspace Gyr.

Our isochronal age estimate for WASP-118 agrees well with the age provided by \citet{Hay16} who used the same approach but using the system parameters derived only from the ground-based observations.

The age discrepancy between the isochronal and gyrochronological age estimate for WASP-107 is significant. Similar discrepancies have been observed for many other K-type stars hosting transiting exoplanets (e.g. \citealt{Maxted15b}). It has been suggested that stars hosting massive short-period planets may have been spun-up by the tidal interaction with the planet and thus exhibit a lower gyrochronological age \citep{Maxted15b}. However, the warm-Saturn WASP-107b is not massive enough and does not orbit close enough to its host star to cause a significant tidal spin-up. A more probable reason for the observed age discrepancy in the case of WASP-107 is the radius anomaly, in which late-type stars exhibit larger radii than is predicted by the stellar models \citep{Popper97}. The radius anomaly is an active research topic driven by the advances in simulating convections in low-mass stars \citep{Ludwig08} and incorporating magnetic fields into stellar models \citep{Feiden13}. \citet{Morales10} have demonstrated that the presence of starspots near the poles of low-mass stars could affect the stellar radii and cause the observed radius anomaly. The age discrepancy of magnetically active K-type WASP-107 may therefore be tentatively attributed to starspots.

\section{CONCLUSIONS}

The two main results presented in this paper are the direct detection of magnetic activity in the short-cadence \textit{K2} lightcurve of WASP-107 and the detection of stellar variability of WASP-118.

The magnetic activity of WASP-107 is manifest firstly as a rotational modulation which gives a stellar rotational period of $17.5\pm1.4$\thinspace d. We also detect a total of 5 definite and 4 possible starspot occultation events. With the planet's orbital period being nearly one-third of the rotational period of the star, we might expect to see the same starspot recurring every three transits. Since we found no evidence of recurring starspots, we suggest that the system is misaligned, unless the starspots' lifetimes are shorter than the rotational period of the star.

The multi-periodic variability in the lightcurve of WASP-118 indicates that the star is likely a low-amplitude $\gamma$\thinspace Doradus pulsator. WASP-118 is a good target for a follow-up asteroseismic analysis in order to obtain more precise stellar parameters and to investigate star--planet interactions as a possible cause for the observed stellar pulsations.

Our refinement of WASP-107 system parameters may also prove beneficial, since the planet lies in the transition region between ice and gas giants. Knowing precise system parameters of such planets is crucial for understanding why some ice giants do not become gas giants.

\section*{ACKNOWLEDGEMENTS}

We thank the anonymous referee for their helpful comments. We also thank Prof. Suzanne Aigrain for her help modifying the K2SC data reduction procedure to accept the \textit{K2} short-cadence data. We gratefully acknowledge the financial support from the Science and Technology Facilities Council (STFC), under grants ST/J001384/1, ST/M001040/1 and ST/M50354X/1. This paper includes data collected by the \textit{K2} mission. Funding for the \textit{K2} mission is provided by the NASA Science Mission directorate. This work made use of PyKE \citep{Still12}, a software package for the reduction and analysis of \textit{Kepler} data. This open source software project is developed and distributed by the NASA Kepler Guest Observer Office. This research has made use of the NASA Exoplanet Archive, which is operated by the California Institute of Technology, under contract with the National Aeronautics and Space Administration under the Exoplanet Exploration Program.

\bibliographystyle{mnras}
\bibliography{bibliography}

\label{lastpage}

\end{document}